\documentclass[preview]{elsarticle}
\usepackage{lineno,hyperref}
\usepackage{tikz}
\usepackage{array}
\usepackage{calc}
\usepackage{setspace}
\usepackage{amsmath}
\usepackage{xspace}
\usepackage{color}
\usepackage{pgfplots}
\usetikzlibrary{shapes.geometric, arrows}

\tikzstyle{startstop} = [rectangle, rounded corners, minimum width=3cm, minimum height=1cm,text centered, draw=black, fill=red!30]
\tikzstyle{io} = [trapezium, trapezium left angle=70, trapezium right angle=110, minimum width=3cm, minimum height=1cm, text centered, draw=black, fill=blue!30, text width = 2cm]
\tikzstyle{process} = [rectangle, minimum width=3cm, minimum height=1cm, text centered, draw=black, fill=orange!30, text width = 2cm]
\tikzstyle{decision} = [diamond, minimum width=3cm, minimum height=1cm, text centered, draw=black, fill=green!30, text width = 2cm]
\tikzstyle{arrow} = [thick,->,>=stealth]
\pgfplotsset{width=10cm,compat=1.9}

\begin{document}

\begin{frontmatter}

\title{XSgen: A Cross Section Generation Methodology for Fuel Cycle Simulation}

\author{Flanagan, Robert}
\address{University of South Carolina}

\author{Scopatz, Anthony}
\address{University of South Carolina}

\begin{abstract}
This work investigates a method for generating medium fidelity reactor model cross sections.
The methodology used here couples a neutron transport code with a depletion
calculator. Together the two can be used to generate time dependent group cross sections
for medium fidelity reactor models in fuel cycle simulation. This work analyzes XSgen, a software
that performs the functions mentioned
above. Addtionally, it is shown that XSgen is capable of creating datasets
for a medium fidelity reactor model known as Bright-lite. The results of this work show
that XSgen produces datasets that match NEA benchmarks for both uranium
based LWR, and MOX reactors.
\end{abstract}


\end{frontmatter}

\section{Introduction}

Reactor simulation is an essential part of nuclear fuel cycle modeling. Modeling strategies
may be categorized in three ways: low, medium, and high fidelity. Each of these modeling
paradigms provides different benefits to a fuel cycle simulation but comes with tradeoffs to
either accuracy, computational effort, or both. The methodology demonstrated in this work, XSgen, 
is a new method for automating the generation of cross section datasets for medium and low fidelity models. 
These datasets use fully described data, allowing users to compare different models under known
conditions and remove any ambiguity that might stem from unknown reactor specification in a 
given dataset. This work is a demonstration of the XSgen methodology and focuses on how this methodology
impacts the cross sections derived from it. 

Accurately generating cross section datasets used in fuel cycle simulation allow for the minimization
of mass flow errors between nuclear facilities, whether they use low or medium fidelity models.
While the work in this discussion does not address, in depth, the impact of XSgen on nuclear fuel cycles 
it is important to mention that this work sets up the basis for using XSgen as a tool for verifying fuel cycle 
simulation.  

Low fidelity models are those models that include limited to no physical computations at run time.
This class of modeling works best for systems that have fixed behaviors. For example, a reactor
during steady state operation may be modeled sufficiently accurately with a low fidelity model.
Such models are often refered to as \emph{recipe-based models}. A recipe represents a predefined
composition. Recipe-based models use fixed input and output compositions to
match fresh and spent
fuel respectively. This lookup table strategy is a computationally efficient mechanism to
represent reactor operation. Alternatively, low fidelity models may include interpolation
techniques to build libraries on the fly. For example, interpolating between recipes for
3\% U-235 enriched fuel, and 4\% enriched fuel might be used to obtain a 3.5\% enriched
fuel recipe.
However, interpolating in this manner fails to directly capture any of the physical changes
that might happen to the core by switching from either 3\% or 4\% enriched fuel (where libraries 
are available) to a 3.5\% fuel (where a library is not available).

Medium fidelity models include physics calculations, but fall short of
full neutron transport or multiphysics calculations. These models typically require
pre-built datasets which are constructed from the results of higher fidelity models.
Medium fidelity models synthesize these pre-built datasets with reasonablly fast physics
algorthims ($<1$ minute execution time). This combination of well structured data and
quick solvers allows for a significatly higher degree of modularity, flexibility, and accuracy
for medium fidelity models over their low fidelity counterparts. Additionally, their average
execution time is orders of magnitude lower than full neutron transport and multiphysics
calculations. Examples of medium fidelity reactor models are CLASS\cite{class},
Bright-lite\cite{brightlite}, and CyBORG\cite{cyborg}.

High fidelity models are constructed using neutronics calculations, depletion
solvers\cite{bateman1910, bateman}, or other coupled multiphysics algorithms. These models tend
to require information on the complete reactor design in order to be executed.
Additionally, due to their higher accuracy, these models require a proportionally heavier amount
of computation. An example of a high fidelity reactor model is the
MCNP\cite{mcnp5monte} burn card which couples Monte Carlo neutron transport with a depletion
calculator.

Medium fidelity models provide a useful middle road between accuracy and computational effort
for nuclear fuel cycle simulators. However, they may vary greatly in the mechanisms and
techniques that they employ. Bright-lite uses a fluence based neutron balance approach to
determine the behavior of the reactor. It also includes optional physics behaviors, such as
disadvantage factors, batch physics, and fuel blending. Alternitavely, CLASS uses a set of
reactor input and output recipes and recursive neural networks\cite{classneural} to dynamically determine
reactor behavior. To understand how these different modeling choices impact a fuel cycle
scenario, it is important for the models to have same fundemental information basis.
At a minimum, the one-group cross sections must be the same.

At present, each medium fidelity model uses its own method for generating input datasets.
This implies that any methodlogical comparison between would be (at least partially) invalid
because each model starts with distinct inputs. Thus, any difference in their outputs
could be attributable to differences in inputs and any similarities in outputs could be
happenstance. Therefore, only an analysis that uses the same baseline datasets
is valid for teasing out differnces that arrise purely from methodology.

This paper discusses a method which generates datasets usable by medium fidelity reactor models.
The authors term this method XSgen and a concrete implementation of this method is available
in the cooresponding BSD-licensed XSgen software package\cite{xsgen}.

XSgen couples a neutron transport code with a depletion solver to automate the generation of
medium fidelity reactor model datasets. XSgen uses OpenMC\cite{openmc} as neutron transport.
A one thousand group, logarithimically spaced flux is extracted from OpenMC. The lower bound for 
this group structure is 1E-9MeV and the upper bound is 10MeV. This flux is
measured in the fuel region of supplied reactor geometry. This flux is fed into a cross
section collapser to generate one-group or multi-group cross sections. This collapse is
performed by PyNE\cite{pyne}. Finally, either the
one-group or multi-group cross sections are coupled to a depletion solver. Currently, this
role is satisfied by ORIGEN v2.2\cite{origen2}, which requires one-group cross sections.

The XSgen method is capable of providing all of the inputs that medium fidelity reactor models
may use: time dependent one- or multi-group cross sections, transmutation matricies,
burnup rates, neutron production rates, and destruction rates.
XSgen stores the data it generates in a single database.
From here, medium fidelity reactor models are able extract the information they need to
derive a new dataset for use in a fuel cycle simulation.

XSgen can also be used to set up more accurate comparisons between low fidelity recipe reactor
models and medium fidelity reactor model types. This is made possible because XSgen may also
simulate a fuel bundle well past its effective lifetime in the core. The data generated by an
XSgen run of this type contains enough information to build out an estimate of core composition
at discharge using reactor batch physics which is discussed in section \S\ref{sec:recipe}.

The following sections will demonstrate the workflow and capabilities of XSgen. The workflow described in section 
\S\ref{sec-workflow} shows 
both the computational work as well as the mathematical basis. Following that is a description of the direct cross 
section results from XSgen in section \S\ref{sec:crosssection}. Finally, XSgen is used for several verification cases with current reactor 
technology in sections \S\ref{sec:LWRv}, \S\ref{sec:startupv}, and \S\ref{sec-mox-benchmark}. 

\section{The XSgen Workflow}
\label{sec-workflow}

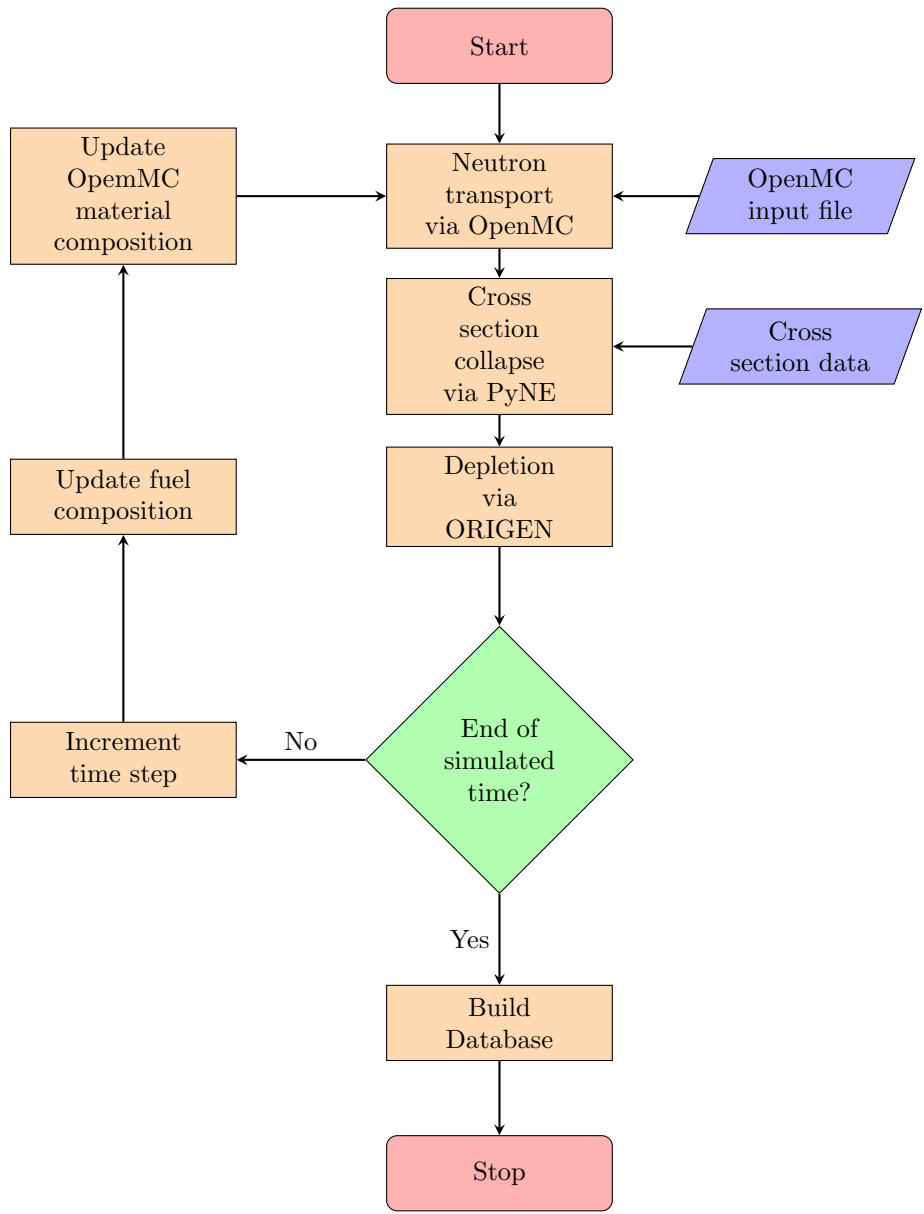
\begin{figure}\center\begin{tikzpicture}[node distance=2cm]
\node (start) [startstop] {Start};
\node (openmc) [process, below of = start]{Neutron transport via OpenMC};
\node (omcinput) [io, right of = openmc, xshift = 2cm]{OpenMC input file};
\node (omcupdate) [process, left of = openmc, xshift = -3cm]{Update OpemMC material composition};
\node (pyne) [process, below of = openmc]{Cross section collapse via PyNE};
\node (pynexs) [io, right of = pyne, xshift = 2cm]{Cross section data};
\node (origen22) [process, below of = pyne]{Depletion via\\ORIGEN};
\node (compupdate) [process, left of = origen22, xshift = -3cm]{Update fuel composition};
\node (timecheck) [decision, below of = origen22, yshift = -1.5cm]{End of simulated time?};
\node (timeinc) [process, left of = timecheck, xshift = -3cm]{Increment time step};
\node (brightlite) [process, below of = timecheck, yshift = -1.5cm]{Build Database};
\node (stop) [startstop, below of = brightlite] {Stop};
\draw [arrow] (start) -- (openmc);
\draw [arrow] (omcinput) -- (openmc);
\draw [arrow] (omcupdate) -- (openmc);
\draw [arrow] (openmc) -- (pyne);
\draw [arrow] (pynexs) -- (pyne);
\draw [arrow] (pyne) -- (origen22);
\draw [arrow] (origen22) -- (timecheck);
\draw [arrow] (compupdate) -- (omcupdate);
\draw [arrow] (timeinc) -- (compupdate);
\draw [arrow] (timecheck) -- node[anchor=south]{No}(timeinc);
\draw [arrow] (timecheck) -- node[anchor=east]{Yes}(brightlite);
\draw [arrow] (brightlite) -- (stop);
\end{tikzpicture}
\caption{Flow chat of the XSgen process for building medium fidelity reactor databases.}
\label{fig:flow}
\end{figure}

The main line of the XSgen workflow, displayed in Figure \ref{fig:flow}, couples three primary tools: neutron transport via OpenMC,
cross section collapse via PyNE, and depletion via ORIGEN v2.2. The first step is the simulation
of the reactor core using a Monte Carlo neutron transport simulation. The primary result of
this is a highly resolved flux spectrum. This spectrum is then collapsed with a standard
cross section database and converted into a suite of one-group cross sections for a reactor
at time step $t$ [days] after the reactor has started up. These one-group cross sections are
then used by a depletion solver to compute the burnup [MWd/kgIHM], neutron production rate [n/s], neutron
destruction rate [n/s], and transmutation matrices. The value of $t$ is set by the user when XSgen
is executed. The composition of the material at the end of the time step is then submitted
back to the neutron transport solver as input. This process is repeated until the maximum time
step is reached. This type of linkage has been made in other softwares, for example LOOP\cite{loop}. However, 
XSgen aims to take the technique further by expanding the physics in the cross section collapse, and providing 
automatic linkage with medium fidelity models.  

\subsection{Neutron Transport via OpenMC}

OpenMC was used as the reference neutron transport calculator for modeling reactors.
It was chosen due to its availablity (it is BSD licensed\cite{open3bsd}), its ability to quickly perform
reactor oriented calculations\cite{serpentvmonte}, and its capability to compute scattering
kernels. Currently, XSgen only obtains the group flux values from OpenMC. It extracts
these for each timestep.

Input templates for OpenMC can be specified in the XSgen run control file.
New templates may be constructed for a reactor design by creating a new input file for OpenMC
and templating it according to a set of field variables that XSgen provides.
XSgen comes stock with standard templates representing pressurized water reactor (PWR)
and fast reactor (FR) lattices. More standard templates may be added to XSgen in the future.

\subsection{Cross Section Collapse via PyNE}

PyNE is used to perform the group collapse
from cross section databases down to the one-group or multi-group cross sections.
PyNE is capable of reading in cross section data from both ACE\cite{ace} and ENDF\cite{endf}
datasets. It is able to synthesize cross section data - that may exist with
many different energy grids - down to a single, standard energy group structure as specified
by the user.

The algorithm PyNE uses for collaspsing cross sections is tailored to efficiently collapsing
the same group structure over many different data sets.
It operates by first constructing a partial energy matrix (PEM). This matrix maps a higher
resolution group structure to a lower resolution group structure. It does this by determining
the contribution of each of the higher resolution energy groups into the lower resolution
energy groups using a weighted sum. A PEM is only applicable for the transformation it is
originally designed for. For example, a PEM that transforms a 10 group system into a
1 group system can not be used to transform a 20 group system into a 1 group system.
The calculations required to generate a PEM can be computationally expensive
and therefore are only performed if a new group structure is added to the system.

\begin{figure}
\caption{Visual representation of the PEM group relations}
\includegraphics[scale=1.2]{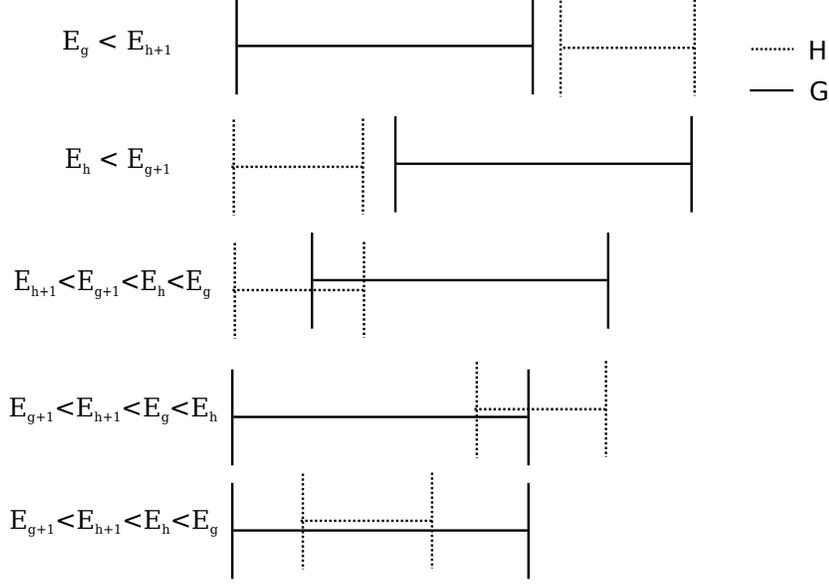}
\label{fig:bin_graph}
\end{figure}

Here, $\vec{\phi_h}[n/s/cm^2]$ represents the high resolution group fluxes for $H$ energy groups with
a group structured defined by $E_h[MeV]$ bin boundaries. Likewise, $\vec{\phi_g}[n/s/cm^2]$ represents
the collapsed flux with $G$ groups and $E_g[MeV]$ bin boundaries. The partial energy
matrix $P$ is defined by the relations seen in Equation \ref{pem-relations}.
Equation \ref{pem-relations} assumes that all energies are monotonically decreasing, namely
that $E_{g+1} \le E_{g} \forall g\in G$ and $E_{h+1} \le E_{h} \forall h\in H$. Figure \ref{fig:bin_graph}
displays the relations shown in Equation \ref{pem-relations}.
\begin{equation}
\label{pem-relations}
P_{g,h} = \left\{
\begin{array}{ll}
    0 & : E_{g} < E_{h+1} \\
    0 & : E_{h} < E_{g+1} \\
    \left(\frac{E_h - E_{g+1}}{E_h - E_{h+1}}\right) & : E_{h+1} \le E_{g+1} \le E_{h} \le E_{g}\\
    \left(\frac{E_{g} - E_{h+1}}{E_h - E_{h+1}}\right) & : E_{g+1} \le E_{h+1} \le E_{g} \le E_{h} \\
    1 & : E_{g+1} \le E_{h+1} \le E_{h} \le E_{g} \\
\end{array}
\right.
\end{equation}

Once $P$ is obtained, collapsing a dataset (such as group fluxes or group constants) from $H$
groups into $G$ groups requires only taking the dot product of the PEM
by the data set. Equation \ref{phi-h-to-g} demonstrates this for group fluxes and Equation
\ref{sig-h-to-g} shows this transition for group constants.
\begin{equation}
\label{phi-h-to-g}
\vec{\phi_g} = P \cdot \vec{\phi_h}
\end{equation}
\begin{equation}
\label{sig-h-to-g}
\vec{\sigma_g} = P \cdot \vec{\sigma_h}
\end{equation}
Importantly, the PEM $P$ may be reused as many times as necessary and the dot product in
Equations \ref{phi-h-to-g} \& \ref{sig-h-to-g} is relatively cheap computationally.

The above PEM expressions are usable when groupwise cross sections are already available.
However, when only continous energy data is available for a nuclide, these pointwise
cross sections must be collapsed to groupwise versions.
In Equation \ref{pointwise-collapse}, $E_p$ represents the energy of a pointwise cross section
in the data set,
$\sigma(E_p)[b]$ represents the cross section at energy $E_p$, and $\sigma_{r,g}^i[b]$ represents a
groupwise cross section for $i^\mathrm{th}$ nuclide and the $r^\mathrm{th}$ reaction channel.
\begin{equation}
\label{pointwise-collapse}
\sigma_{r,g}^i = \frac{1}{2} \frac{\sum_{E_p\ge E_{g+1}}^{E_p\le E_g}
                                   \left(\sigma_r^i(E_p)+\sigma_r^i(E_{p+1})\right)
                                   \left(E_{p+1}-E_{p}\right)}
                                  {E_g-E_{g+1}}
\end{equation}
Equation \ref{pointwise-collapse} represents a trapezoidal integration weighted by the
energy width of the group. From here, one-group cross sections may be obtained using the PEM
matrix from above. Note that in Equation \ref{pointwise-pem}, $\phi$ represents the total
flux ($\phi=\sum\phi_g$).
\begin{equation}
\label{pointwise-pem}
\sigma_{r}^i=\frac{P \cdot (\overrightarrow{\sigma_{r,g}^i \phi_g})}{\phi}
\end{equation}
Here $\sigma_{r}^i$ represents the one-group cross section of the $r^\mathrm{th}$ reaction
channel for the $i^\mathrm{th}$ nuclide, and $\overrightarrow{\sigma_{r,g}^i \phi_g}$ is the
elementwise multplication of the group constants and the group fluxes.
This process is repeated for all nuclides and reaction channels. The resultant one-group
cross section dataset is used to construct a custom ORIGEN v2.2 TAPE9 file\cite{origen2}
for each time step.

PyNE also enables the user to incorperate self-shielding effects \cite{bondarenko} into the cross sections
computed for XSgen. It does so by building an energy dependent function of weights for each
nuclide. These weights scale the one-group cross section depending on the density of the
nuclide. The more concentrated a nuclide is within a material composition, the lower the
effective cross section is of that nuclide. Self-shielding weights are computed by the
expression in Equation \ref{ssw} \cite{weights}.
\begin{equation}
\label{ssw}
\omega_{g}^i=\left(\frac{\sum_{j\neq i}N_j \, \sigma_{t,g}^j}{N_i}+\sigma_{t,g}^i\right)^{-1}
\end{equation}
In this expression, $I$ represents the set of all nuclides, and $i$ and $j$ index $I$.
 $N_j$ represents the number density [1/cm$^3$] of
nuclide $j$, and $\sigma_{t,g}^i$ is the flux weighted total cross section of
the $i^\mathrm{th}$ nuclide in energy bin $g$.

\begin{figure}[h]
  \center
  \includegraphics[scale=0.6]{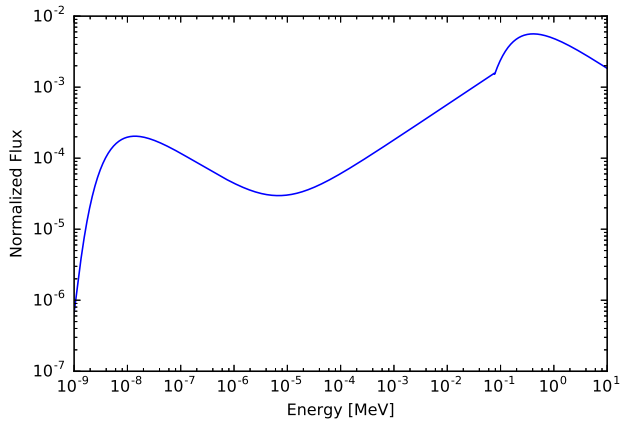}
  \caption{Representative pointwise flux for PWR the caclulation of self-shielding weights}
  \label{fig:therm}
\end{figure}

The total cross sections in Equation \ref{ssw} may be computed from continuous energy
pointwise ENDF or ACE formatted data for those species where such data sets are available.
The method for computing the  $\sigma_{t,g}^i$ is the same as in Equation
\ref{pointwise-collapse} \& \ref{pointwise-pem} for a single energy group.  However,
the challenge in this circumstance is obtaining a pointwise flux. OpenMC does not output
a pointwise flux, so a surrogate pointwise flux is used which approximates the normalized shape
of the reactor. The equation for a surrogate pointwise flux for a PWR may be seen in
Equation \ref{pointwise-flux} \cite{spectrum}.
\begin{equation}
\label{pointwise-flux}
\phi =
\begin{cases}
    \frac{2\pi\sqrt{2E*1e6}}{(\pi kT)^{3/2}}e^{\frac{-E*1e6}{kT}} & : E\leq 0.155\times 10^{-6} \\
    \frac{1}{\sqrt{2E*1e6}} + 0.453e^{-1.036E}\cdot\sinh{\sqrt{2.29E}} & : E > 0.155\times 10^{-6} \\
\end{cases}
\end{equation}
Here, $k$ is the Boltzman constant $[\frac{eV}{K}]$, $T$ is the temperature [K],
and E is the energy [MeV]. Equation \ref{pointwise-flux} is displayed in
Figure \ref{fig:therm}.

Thus, Equation \ref{ssw} represents a two-dimensional matrix of dimensions
$I$ (number of nuclides) by $G$ (number of groups). This entire matrix 
must be recalculated every time step. This is because the number densities of the nuclides
in the fuel will charge as burnup increases, invalidating any previously obtained weights.

Finally these weights are applied to equation \ref{pointwise-pem}:
\begin{equation}
\label{ssww}
\sigma_{r}^i=\frac{P \cdot (\overrightarrow{\omega_g^i\sigma_{r,g}^i \phi_g})}{\phi}
\end{equation}

In this case $\phi$ is also calculated using the weights:
\begin{equation}
\label{phiw}
\phi=\sum\phi_g \cdot \omega_g^i
\end{equation}

\begin{figure}[h]
  \center
  \includegraphics[scale=0.6]{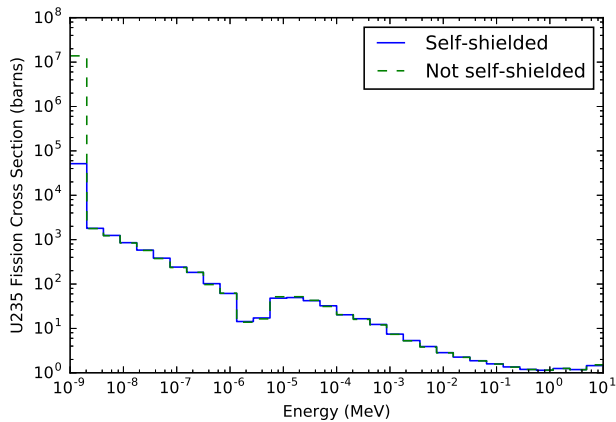}
  \caption{The effects of self-shielding on the fission cross section of Uranium-235 inside a fuel pellet at 3.5\%.}
  \label{fig:index}
\end{figure}

The effects of self shielding on U-235 inside of a fuel pellet man be seen in
Figure \ref{fig:index}. By inspection, the self-shielding effect can be large - and thus
necessary to include - for resonance cross sections and low energy capture cross sections.

\subsection{Depletion via ORIGEN v2.2}
\label{sec:origen}

After a TAPE9 file (the cross section format of ORIGEN v2.2) has been constructed by XSgen, it is
used to perform two types of burnup calculations. The first is performed on the fuel and
includes the entire fuel material composition. Once this is done, the material at the
end of the fuel burn is passed back to OpenMC to start the next XSgen time step iteration.

The second type of depletion calculation is used to build a transmutation dataset for a
specific nuclide. These runs take one kilogram of a pure, intial nuclide as the input to
ORIGEN. At the end of the depeltion, the burnup, neutron production rate, neutron
destruction rate, and transmutation matrix are all recorded. The nuclides of interest
for which these parameters are tracked are chosen ahead of time by the XSgen user.
This nuclide set represents which nuclides are admissable as fresh fuel for the medium
fidelity reactor model.

\subsection{XSgen as a Recipe Generator}
\label{sec:recipe}
XSgen may be used to construct recipes for low fidelity reactor models. Recipes creation
requires simulating the full reactor lifetime. One technique is to use a high fidelity
model to accurately model the burnup and transmutation of a reactor core from loading to
discharge of a fuel assembly. XSgen performs precisely this calculation.

XSgen necessarily tracks a single kilogram of fresh fuel through the reactor. When this
fuel becomes subcritical it represents a single batch reactor core becoming subcritical.
For a single batch core, this is when refueling would be required. However, it is possible
to predict the burnup of a multi-batch core from the burnup of a single batch core.
Following the linear reactivity model \cite{linear}, Equation \ref{burnup-n} estimates the burnup
$BU_n$ [MWd/kg] of an $n$-batch core based on the burnup of a single batch core, as seen in equation \ref{burnup-n}.
\begin{equation}
\label{burnup-n}
BU_n = \frac{2n}{n+1} \, BU_1
\end{equation}
This technique works primarily with reactor cores that have a solid fuel which undergoes
fuel shuffling during a refueling operation. Using the predicted burnup of the fuel,
it is possible to extract the fluence, and therefore output composition,
of the fuel at that burnup\cite{brightlite, linear}. 

Note that this technique assumes that all batches are independent of one another
Additionally, Equation \ref{burnup-n} works best for reactors that have a decreasing
criticality over a cycle. Specifically, it does not provide valid recipes for accelerator
driven systems.

\section{Cross Section Generation Results}
\label{sec:crosssection}
\begin{figure}
\includegraphics[scale=0.8]{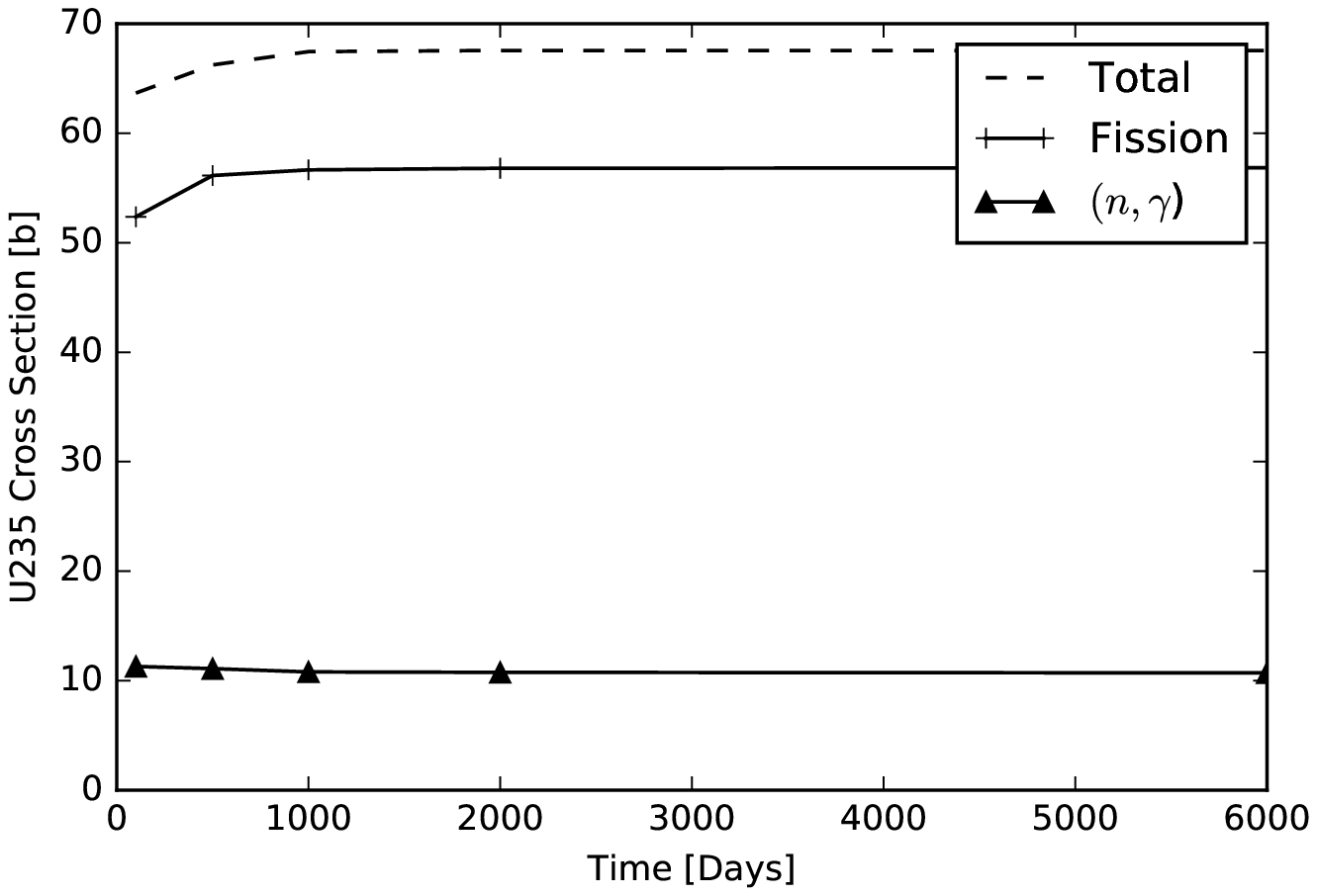}
\caption{Cross Section Behavior of U235 in 3.5\% enriched LWR.
         The U235 is subjected to a constant flux of $3\times10^{14}[n/s/cm^2]$.}
\label{fig:u235xs}
\end{figure}

\begin{figure}
\includegraphics[scale=0.8]{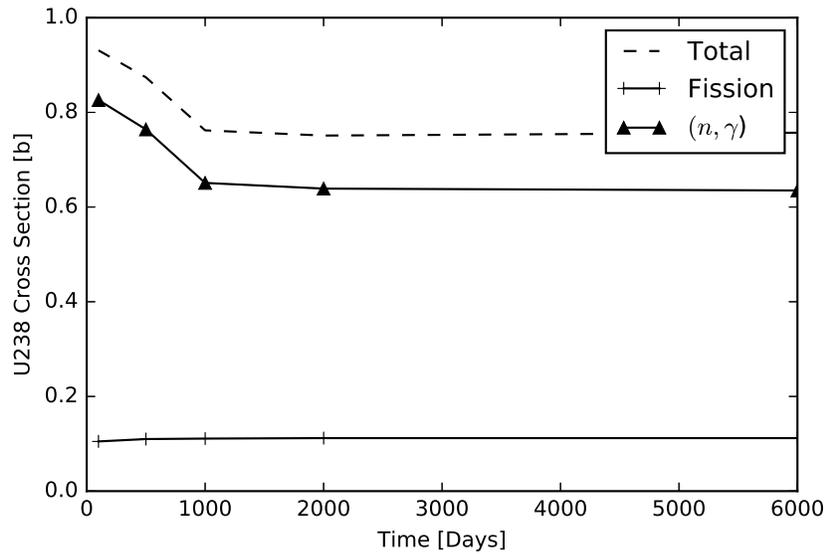}
\caption{Cross Section Behavior of U238 in 3.5\% enriched LWR. The U238 is subjected to a constant flux of $3\times10^{14}[n/s/cm^2]$.}
\label{fig:u238xs}
\end{figure}

To test the capability of XSgen to generate cross sections, a test case was run emulating a LWR. 
Figures \ref{fig:u235xs} and \ref{fig:u238xs} show cross section behavior results
from running XSgen for the LWR using the input parameters found in
Table \ref{tab:xsgenstats1}. The fresh fuel used was 3.5\% enriched uranium oxide (UOX).
The number of energy groups used for this analysis was 1000, log-uniformly spaced between
0.001 eV and 10 MeV. This test was conducted using the pin-cell model\cite{pin-cell}. 

\begin{table}[!htb]
\centering
\caption{Reactor Input Parameters for XSgen Comparisons}
\label{tab:xsgenstats1}
\begin{tabular}{ll}
Input & Value \\
\hline
Fuel Cell Radius [cm] & 0.410 \\
Void Cell Radius [cm] & 0.4185 \\
Clad Cell Radius [cm] & 0.475 \\
Unit Cell Pitch  [cm] & 1.32 \\
Unit Cell Height [cm] & 10.0 \\
Fuel Density [g/cc] & 10.7 \\
Clad Density [g/cc] & 5.87 \\
Coolant Density [g/cc] & 0.73 \\
\end{tabular}
\end{table}

Figures \ref{fig:u235xs} \& \ref{fig:u238xs} show how the the fission and (n, $\gamma$)
cross sections change slowly over the course of operating the reactor. This is to be
expected as the flux and number densities of U235 and U238 change. 
This behavior is consistent with the expected behavior of cross sections within an LWR.

\begin{figure}[h]
  \center
  \includegraphics[scale=0.7]{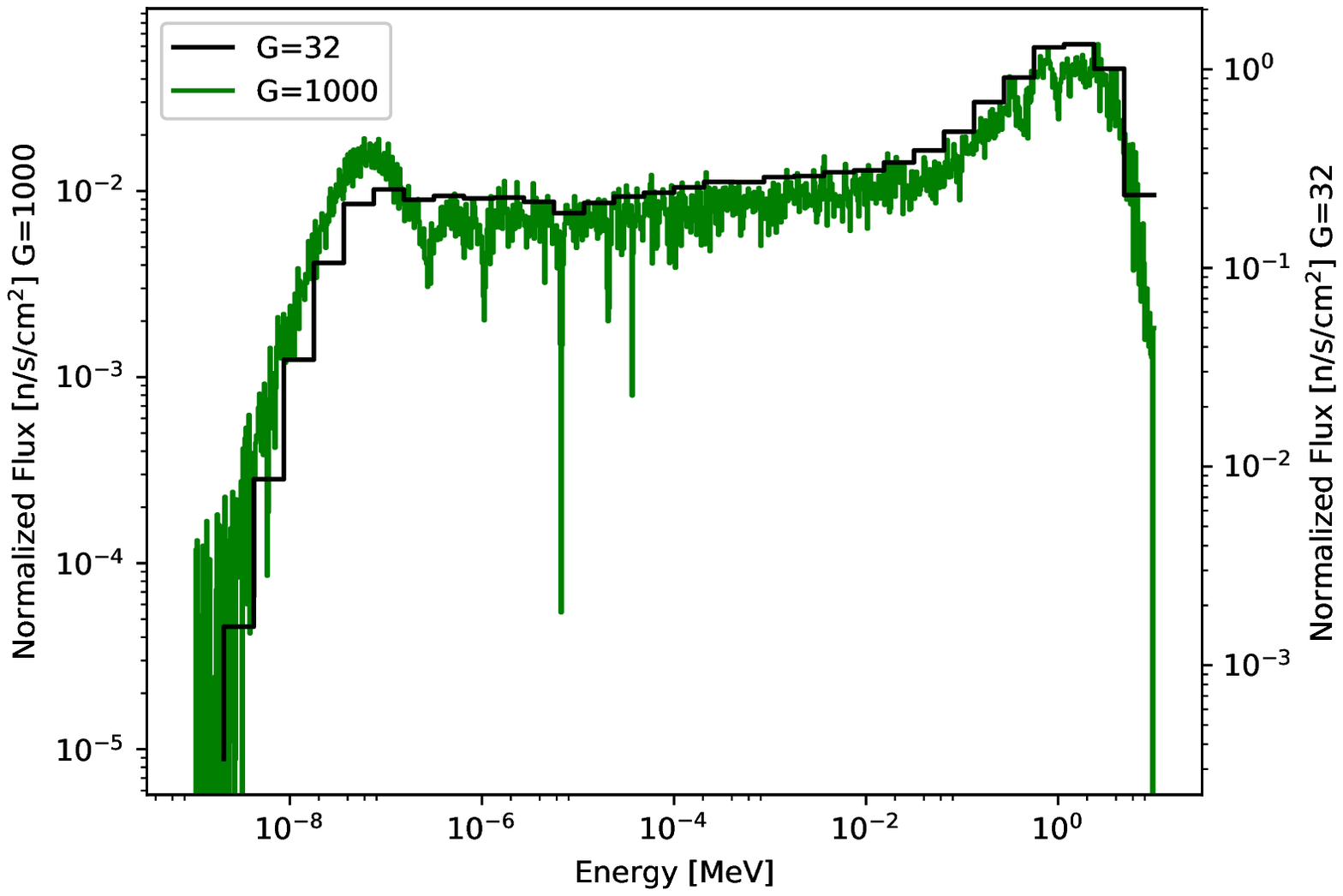}
  \caption{Fluxes generated by OpenMC for a light water reactor using 32 and 1000 group energy structures.}
  \label{fig:32g}
\end{figure}

Moreover, the following figure shows the equilibrium multi-group flux of the reactors, as
determined by OpenMC.
Figure \ref{fig:32g} shows the fluxes of the same XSgen run using two different energy group
structures; namely 32 \& log-uniformly spaced groups.
Figure \ref{fig:32g} demonstrated the thermal spectrum behavior of the two group structures.
The larger, 1000-group structure manages to catch the effects of stronger resonance groups
as well as providing better detail than the lower, 32-group structures.

\section{LWR Verification Cases}
\label{sec:LWRv}
XSgen will be verified in conjunction with Bright-lite to produce burnups and
compositions for several known reactor systems. As Bright-lite has already been tested
and verified with other input datasets\cite{brightlite},
any deviations within the results come from the XSgen method.

The cases presented here represent a range of different enrichments in several LWRs.
Results from XSgen/Bright-lite are compared to recipes provided with VISION\cite{vision}.
Specifically, cases at 3.0\%, 3.5\%, and 4.0\% U-235 enrichment where tested.
Three batch cores were assumed for all the enrichment test cases.

The results presented in Tables \ref{tab:30}-\ref{tab:40}
all compare used fuel composition using mass fractions and percent difference.For this 
calculation VISION is used as a reference set of values. 
 
\begin{equation}
\label{percentdiff}
\%\mathrm{ Diff} = \mathrm{\frac{value_{accepted} - value_{tested}}{value_{accepted}}\times100}
\end{equation}

\begin{table}[!htb]
\centering
\small
\caption{The output nuclides at equalibrium for VISION and Bright-lite at 3.0\% enriched UOX. Difference is displayed as \% difference}
\label{tab:30}
\vspace{0.5em}
\begin{tabular}{cccc}
Nuclide &  VISION & Bright-lite & \% Diff \\
\hline
U235 & 6.75E-3 & 6.70E-3 & -0.78\%\\
U238 & 9.45E-1 & 9.41E-1 & -0.42\%\\
Pu238 & 1.23E-4 & 1.27E-4 & 3.01\%\\
Pu239 & 5.15E-3 & 5.25E-3 & 1.95\%\\
Pu240 & 2.38E-3 & 2.40E-3 & 0.97\%\\
Pu241 & 1.30E-3 & 1.35E-3 & 3.69\%\\
Pu242 & 5.43E-4 & 5.48E-4 & 0.88\%\\
Am241 & 3.56E-5 & 3.86E-5 & 8.81\%\\
Am243 & 9.38E-5 & 9.01E-5 & -3.94\%\\
Cm242 & 1.38E-5 & 1.34E-5 & -3.19\%\\
Cm244 & 2.79E-5 & 2.57E-5 & -7.94\%\\
\end{tabular}
\end{table}

\begin{table}[!htb]
\centering
\small
\caption{The output nuclides at equalibrium for VISION and Bright-lite at 3.5\% enriched UOX. Difference is displayed as \% difference}
\label{tab:35}
\vspace{0.5em}
\begin{tabular}{cccc}
Nuclide &  VISION & Bright-lite & \% Diff \\
\hline
U235 & 6.80E-3 & 6.71E-3 & -1.30\%\\
U238 & 9.35E-1 & 9.30E-1 & -0.53\%\\
Pu238 & 1.85E-4 & 1.86E-4 & -0.41\%\\
Pu239 & 5.41E-3 & 5.45E-3 & 0.66\%\\
Pu240 & 2.60E-3 & 2.61E-3 & 0.25 \%\\
Pu241 & 1.48E-3 & 1.52E-3 & 2.74\%\\
Pu242 & 6.92E-4 & 6.93E-4 & 0.10 \%\\
Am241 & 4.59E-5 & 5.17E-5 & 12.60\%\\
Am243 & 1.37E- & 1.26E-4 & -7.47\%\\
Cm242 & 1.88E-5 & 1.79E-5 & -4.91\%\\
Cm244 & 4.38E-5 & 4.54E-5 & -6.06\%\\
\end{tabular}
\end{table}

\begin{table}[!htb]
\centering
\small
\caption{The output nuclides at equalibrium for VISION and Bright-lite at 4.0\% enriched UOX. Difference is displayed as \% difference}
\label{tab:40}
\vspace{0.5em}
\begin{tabular}{cccc}
Nuclide & VISION & Bright-lite & \% Difference \\
\hline
U235 & 6.95E-3 & 6.91E-3 & -0.56\%\\
U238 & 9.27E-1 & 9.24E-1 & -0.32\%\\
Pu238 & 2.52E-4 & 2.55E-4 & 1.35\%\\
Pu239 & 5.64E-3 & 5.67E-3 & 0.57\%\\
Pu240 & 2.78E-3 & 2.79E-3 & 0.4\%\\
Pu241 & 1.62E-3  & 1.62E-3 & -0.02\%\\
Pu242 & 8.17E-4  & 8.19E-4 & -0.28\%\\
Am241 & 5.55E-5 & 6.04E-5 & 8.79\%\\
Am243 & 1.76E-4 & 1.74E-4 & -1.40\%\\
Cm242 & 2.33E-5 & 2.31E-5 & -0.83\%\\
Cm244 & 7.06E-5 & 7.00E-5 & -0.90\%\\
\end{tabular}
\end{table}

Tables \ref{tab:30}-\ref{tab:40} shows a very strong agreement between VISION and Bright-lite
results using datasets created by XSgen from light water reactor designs using these particular
enrichments. Bright-lite has been benchmarked against VISION previously using other datasets
\cite{brightlite}. The previously benchmarked Bright-lite used stock Origen2.2 libraries\cite{origen2}. This shows that
XSgen generated datasets allow for Bright-lite to match VISION to a 5\% tolence for a
majority of nuclides. Those that exhibit the greatest deviations are high order transuranics. U238 shows the least
deviation for all nuclides. Since the higher order actinides are generated from U238, the error in U238
is an indication that the overall error of the system is low. 

Am241 shows the largest deviations from the VISION data. There is more Am241 in every
case than in the VISION recipes. There is thus less of the species that are arise due to
the presence of Am241, namely Am243, Cm242, and Cm244. These results can be explained by a
neutron absorption cross section error in Am241. If the neutron absorption cross section for
Am241 is too low, it can never transmute into Am242 and Am243. This in turn leads to a
relative sparsity of curium nuclides.

\section{LWR Startup Behavior Verification}
\label{sec:startupv}
The Bright-lite model generated by XSgen was additionally tested against a set of
Nuclear Enegy Agency (NEA) results\cite{nea}. These results demonstrate the start up behavior
of the same reaction using two different fuel enrichments.

Table \ref{tab:b} shows how XSgen/Bright-lite compares to the NEA results.
The equilibrium results here show good agreement with this case. As Bright-lite is often
used to derive output compositions from used fuels in recycle scenarios, accurately
predicting the equilibrium discharge is important feature of medium fidelity reactor model
input datasets.

\begin{table}[!htb]
\centering
\caption{NEA core discharge data for a 3.1\% enriched light water reactor with startup behavior.}
\label{tab:b}
\begin{tabular}{lllll}
Batch & Burnup (MWd/kg) & U235 (w\%) & Fissile Pu (w\%) & Total Pu(w\%) \\
\hline
1 & 12.04 & 0.64 & 0.464 & 0.633 \\
2 & 23.86 & 0.76 & 0.6 & 0.818 \\
3 & 31.75 & 0.8 & 0.677 & 0.921 \\
4 & 32.00 & 0.85 & 0.697 & 0.943 \\
5+ & 33.00 & 0.85 & 0.688 & 0.943
\end{tabular}
\end{table}

\begin{table}[!htb]
\centering
\caption{The startup values and percent difference from the NEA data for the XSgen/Bright-lite reactor system.}
\label{tab:c}
\begin{tabular}{l| ll| ll| ll| ll}
 & \multicolumn{2}{p{1cm}}{Burnup [MWd/kgIHM]} & \multicolumn{2}{c}{U235 [w\%]} & \multicolumn{2}{c}{Fissile Pu [w\%]} & \multicolumn{2}{c}{Total Pu [w\%]} \\
Batch & Value & \%Diff & Value & \%Diff & Value & \%Diff & Value & \%Diff \\
\hline
1 & 13.570 & 12.754 & 0.656 & 2.50 & 0.550 & 18.737 & 0.680 & 7.541 \\
2 & 22.040 & -7.628 & 0.723 & -4.868 & 0.641 & 6.919 & 0.835 & 2.189 \\
3 & 32.510 & 2.394 & 0.789 & -1.375 & 0.710 & 4.883 & 0.970 & 5.231 \\
4 & 31.230 & -2.406 & 0.851 & 0.117 & 0.680 & -2.254 & 0.906 & -5.031 \\
5 & 33.010 & 0.030 & 0.843 & -0.824 & 0.703 & 2.198 & 0.946 & 0.285 \\
6+ & 33.020 & 0.060 & 0.855 & 0.588 & 0.700 & 1.833 & 0.951 & 0.820
\end{tabular}
\end{table}

The results of the startup nuclide loading can be seen in Table \ref{tab:c}.
The first discrepency in these results is the amount of plutonium generated
within the first batch is significantly higher within Bright-lite than in the NEA data.
This trend only continues for the first two cycles before settling out. In the NEA data, the start up batches 
are placed around the core optimally, with lower enrichment fuel bundles blended throughout the 
higher enrichment fuel bundles. In Bright-lite the fuel bundles are all grouped, meaning the center of 
the reactor is primarily comprised of U238 (98.8\% U238) fuel bundles. Higher U238 content
increases the production of plutonium in the first few batches. Until the batches reaching the center of
the reactor have a lower U238 content and fission products that reduce the U238 transmutation 
rate the nuclide vectors will not agree well. The high amount of U238 in these initial batches causes an elevated production rate of
Pu239 over the NEA data. The NEA benchmark has these low enrichment startup batches
spread more evenly. Meanwhile, Bright-lite simulates them as being the inner most section of the core.

\begin{table}[!htb]
\centering
\caption{NEA data for a 3.6\% enriched light water reactor start up behavior.}
\label{tab:d}
\begin{tabular}{lllll}
Batch & Burnup (MWd/kg) & U235 (w\%) & Fissile Pu (w\%) & Total Pu(w\%) \\
\hline
1 & 13.9 & 0.840 & 0.474 & 0.629 \\
2 & 22.67 & 0.721 & 0.642 & 0.892 \\
3 & 32.36 & 0.647 & 0.716 & 1.039 \\
4 & 41.00 & 0.640 & 0.785 & 1.177 \\
5 & 39.00 & 0.940 & 0.808 & 1.166 \\
6 & 40.60 & 0.88 & 0.817 & 1.194 \\
7+ & 42.50 & 0.81 & 0.827 & 1.223
\end{tabular}
\end{table}

\begin{table}[!htb]
\centering
\caption{XSgen / Bright-lite data for the 3.6\% enriched light water reactor start up behavior.}
\label{tab:e}
\begin{tabular}{l | ll | ll | ll | ll}
 & \multicolumn{2}{p{1cm}}{Burnup [MWd/kgIHM]} & \multicolumn{2}{c}{U235 [w\%]} & \multicolumn{2}{c}{Fissile Pu [w\%]} & \multicolumn{2}{c}{Total Pu [w\%]} \\
Batch & Value & \%Diff & Value & \%Diff & Value & \%Diff & Value & \%Diff \\
\hline
1 & 13.37 & -3.84 & 0.930 & 10.77 & 0.62 & 30.43 & 0.78 & 23.53 \\
2 & 22.69 & 0.07 & 0.82 & 13.02 & 0.72 & 12.81 & 0.96 & 7.49 \\
3 & 32.38 & 0.03 & 0.72 & 11.08 & 0.78 & 8.46 & 1.07 & 2.69 \\
4 & 42.57 & 3.83 & 0.61 & -5.33 & 0.804 & 2.42 & 1.14 & -2.94 \\
5 & 41.00 & 5.13 & 0.85 & 9.526 & 0.79 & -2.81 & 1.10 & -5.91 \\
6 & 43.83 & 3.04 & 0.82 & 6.53 & 0.79 & -3.58 & 1.11 & -7.45 \\
7+ & 42.32 & -0.43 & 0.81 & -0.44 & 0.79 & -4.54 & 1.11 & -9.42
\end{tabular}
\end{table}

Table \ref{tab:d} and Table \ref{tab:e} are analogous to Table \ref{tab:b} and Table \ref{tab:c}
for the 3.6\% U235 enriched LWR case. Again, a similar trend is visible in this case with the
exception that the equilibrium results for the total amount of plutonium within the
reactor is 9.2\% lower in the Bright-lite model. The amount of fissile plutonium at equalibrium is also lower,
but within 5\%. The lower amount of fissile plutonium leads
to less build up of the high order actinides.

\section{MOX Verification}
\label{sec-mox-benchmark}
To demonstrate the full capability of XSgen, the method must be applied to more advanced
reactor types than base case LWRs. To this end, XSgen was used to generate a mixed-oxide (MOX)
fuel library for Bright-lite. This was then applied to a single pass MOX fuel cycle scenario.
In order to maintain a valid comparison, the input fuel composition into the Bright-lite MOX
reactor was exactly the same as a VISION MOX reactor.
This test demonstrates the flexibility of the combined XSgen/Bright-lite system, which enables
the modeling of advanced reactor types with much reduced user effort.

\begin{table}[!htb]
\centering
\caption{Comparison of VISION and Bright-lite output from a single pass MOX reactor.}
\label{tab:g}
\begin{tabular}{lllll}
Nuclide & Input Composition & VISION & Bright-lite & \%Diff \\
\hline
U234 & 2.20E-4 & 2.11E-4 & 2.00E-4 & -5.0\\
U235 & 7.08E-3 & 4.05E-3 & 4.00E-3 & -1.2\\
U236 & 5.28E-3 & 4.97E-3 & 5.60E-3 & 12.6\\
U238 & 8.80E-1 & 8.51E-1 & 8.62E-1 & 1.3\\
Pu238 & 2.85E-3 & 3.22E-3 & 3.01E-3 & -6.5\\
Pu239 & 5.66E-2 & 3.31E-2 & 3.15E-2 & -4.6\\
Pu240 & 2.70E-2 & 2.42E-2 & 2.54E-2 & 5.0\\
Pu241 & 1.17E-2 & 1.31E-2 & 1.27E-2 & -3.3\\
Pu242 & 8.00E-3 & 8.90E-3 & 8.68E-3 & -2.6\\
Am241 & 1.18E-3 & 1.72E-3 & 1.60E-3 & -7.2\\
Am243 & 0.0 & 1.96E-3 & 1.83E-3 & -6.6\\
Cm242 & 0.0 & 2.62E-4 & 2.46E-4 & -6.4\\
Cm244 & 0.0 & 1.03E-3 & 9.84E-4 & -4.4
\end{tabular}
\end{table}

The data presented in Table \ref{tab:g} displays the results of Bright-lite given the XSgen MOX
reactor cross sections. Overall, the Bright-lite model obtains output compositions that
are quite similar to the output recipe of VISION. For most actinides listed, the relative error
is less than 5\%, even for higher order species such as the curium nuclides.
The differences with respect to some of the higher order actinides (Am241 and higher) most
likely stem from the relatively high amount of Pu240. This suggests that Pu240 neutron capture
cross section is lower in Bright-lite. The low Pu240 capture cross section thus lowers the equilibrium concentration of Pu241 and
nuclides that derive from it. A lower cross section could arise from a difference in the cross
section data sets used or in methodology; the pointwise flux used in the self shielding calculation could
be different, the cross section data could be different, or similar input data issues. This highlights exactly 
why XSgen was developed. Unfortunately in this case without the background data on the VISION recipes, XSgen 
could not remove these input data issues. 

Another potential cause of composition uncertainity is that the exact dimensions and behavior
of the VISION MOX reactor are not availble in the VSION database. In order to get the
most accurate cross sections and compositions, XSgen must hav a full and complete reactor
specification. Since such a specification was not available for this MOX case,
a generic 17x17 LWR core was chosen instead. Thus the descrepencies seen in the output
compositions in Table \ref{tab:g} are reasonable. Even without knowing the
true MOX reactor used to formulate the VISION recipes, the resultant compositions line
up well enough for the neutronically important nuclides.

\section{Recipe Reactor Generation}

As discussed in \S\ref{sec-workflow}, XSgen may be used to create recipe reactors.
This section displays a comparison between VISION and XSgen recipe reactors. These LWRs accept
a 3.2\% U235 enriched fresh fuel. The XSgen model was constructed with reactor design
information found in Table \ref{tab:xsgenstats}.

\begin{table}[!htb]
\centering
\caption{Specification forthe XSgen 33 MWd/kgIHM Burnup LWR.}
\label{tab:xsgenstats}
\begin{tabular}{ll}
Input & Value \\
\hline
Fuel Cell Radius [cm] & 0.410 \\
Void Cell Radius [cm] & 0.4185 \\
clad Cell Radius [cm] & 0.475 \\
Unit Cell Pitch  [cm] & 1.32 \\
Unit Cell Height [cm] & 10.0 \\
Fuel Density [g/cc] & 10.7 \\
Clad Density [g/cc] & 5.87 \\
Coolant Density [g/cc] & 0.73 \\
\end{tabular}
\end{table}

Table \ref{tab:recipe} shows the difference between the VISION 3.2\% LWR case and the
cooresponding XSgen reactor discharge recipes.

\begin{table}[!htb]
\centering
\caption{Comparison of VISION and XSgen for the generation of a 33 MWd/kgIHM Burnup.}
\label{tab:recipe}
\begin{tabular}{llll}
Nuclide & VISION & XSgen & \%Diff \\
\hline
U235 & 8.06E-03 & 7.84E-03 & -2.85 \\
U238 & 9.44E-01 & 9.67E-01 & 2.36 \\
Pu238 & 1.09E-04 & 1.01E-04 & -7.80 \\
Pu239 & 5.13E-03 & 4.83E-03 & -6.15 \\
Pu240 & 2.25E-03 & 2.41E-03 & 6.52 \\
Pu241 & 1.22E-03 & 1.17E-03 & -4.53 \\
Pu242 & 4.73E-04 & 4.49E-04 & -5.26 \\
Am241 & 2.98E-05 & 2.76E-05 & -7.97 \\
Am243 & 7.90E-05 & 7.23E-05 & -9.31 \\
Cm242 & 1.17E-05 & 1.07E-05 & -9.48 \\
Cm244 & 2.22E-05 & 2.11E-05 & -5.11 \\
\end{tabular}
\end{table}

The results in Table \ref{tab:recipe} show that XSgen is with 10\% error on all output
compositions for this reactor design. A possible cause for this differnce is the
method used for extrapolating the data for a multi-batch core from a single batch core using
the linear reactivity model. Additionally, the design of the VISION reactor is unknown, similar to the MOX case presented
in \S\ref{sec-mox-benchmark}. The specifications - geometry, material densities, temperature, etc - used for the VISION reactor
are unknown in this case. Output nuclides will vary with respect to those specifications mentioned above. Therefore while
the two cases share a common enrichment and burnup, their exact output compositions will vary. 

This result can be used as a starting point to further improve XSgen recipe reactor generation capabilities. A more detailed 
description of the VISION reactor is required to improve upon these results within XSgen. Additionally, alternative methods 
beyond the linear reactivity model\cite{linear} may provide future improvements.

\section{Conclusion}

The work of nuclear reactor models is to provide designers and researchers with insight
into the behavior of reactors, and by extension nuclear fuel cycles. Comparisons between
these models is important to understanding how the choice of model might impact the
accuracy of the results. This work has demonstrated that XSgen is capable of
providing one-group cross sections for a wide variety of reactor types.
Producing a cross section data set using the exact same specifications for two different reactor models 
can be done automatically. Because equitable comparisons are now possible, the XSgen tool allows
researchers to understand how different reactor models affect the results of fuel cycle analysis.

Additionally, XSgen provides a method for automating the generation of reactor types for
medium fidelity models. The work here linking XSgen to the Bright-lite reactor model shows
that XSgen can be used to expand the set of reactor types and reactor designs
(i.e. varied burnups, enrichment, or core structure) that Bright-lite can represent.
Through this coupling, Bright-lite can be used to quickly examine fuel cycles with new or
interesting reactor technologies.

Future work aims to expand this coupling to include other medium fidelity models such as CLASS.
This will allow for comparisons between medium fidelity models to be performed using the
same source for datasets. Moreover, XSgen currently does not have a stock template for
fast reactor designs or accelerator driven systems. Specialized reactor templates may be
needed for future fuel cycle analysis campaigns.

\section{Acknowledgements}
The preparers of this work would like to thank Dr. Erich Schneider, and Dr. Cem Bagdatlioglu for their
support. Their efforts provided the researchers with the knowledge on how to link the XSgen methodology
to the Bright-lite infrustructure. Additional thanks to the Nuclear Engineering University Program for
funding this work under project number 12-4065.

\bibliography{XSgen}

\begin{thebibliography}{10}
\expandafter\ifx\csname url\endcsname\relax
  \def\url#1{\texttt{#1}}\fi
\expandafter\ifx\csname urlprefix\endcsname\relax\def\urlprefix{URL }\fi
\expandafter\ifx\csname href\endcsname\relax
  \def\href#1#2{#2} \def\path#1{#1}\fi

\bibitem{class}
M.~Baptiste, C.~Jean-Baptiste, T.~Nicolas, Class, a new tool for nuclear
  scenarios: Description \& first application, International Journal of
  Engineering and Physical Sciences 6~(52) (2012) 65.

\bibitem{brightlite}
C.~Bagdatlioglu, E.~Schneider, Method for accounting for macroscopic
  heterogeneities in reactor material balance generation in fuel cycle
  simulations, Nuclear Engineering and Design 302 (2016) 37--45.

\bibitem{cyborg}
S.~E. Skutnik, N.~C. Sly, J.~L. Littell, Cyborg: An origen-based reactor
  analysis module for cyclus, in: Transactions of the American Nuclear Society,
  Vol. 115, 2016, pp. 299--301.

\bibitem{bateman1910}
H.~Bateman, The solution of a system of differential equations occurring in the
  theory of radioactive transformations, in: Proc. Cambridge Philos. Soc,
  Vol.~15, 1910, pp. 423--427.

\bibitem{bateman}
J.~Cetnar, General solution of bateman equations for nuclear transmutations,
  Annals of Nuclear Energy 33~(7) (2006) 640--645.

\bibitem{mcnp5monte}
X.~MCNP, Monte carlo team, mcnp�a general purpose monte carlo n-particle
  transport code, version 5, Tech. rep., LA-UR-03-1987, Los Alamos National
  Laboratory, April 2003. The MCNP5 code can be obtained from the Radiation
  Safety Information Computational Center (RSICC), PO Box 2008, Oak Ridge, TN,
  37831-6362 (5).

\bibitem{classneural}
B.~Leniau, B.~Mouginot, N.~Thiolliere, X.~Doligez, A.~Bidaud, F.~Courtin,
  M.~Ernoult, S.~David, A neural network approach for burn-up calculation and
  its application to the dynamic fuel cycle code class, Annals of Nuclear
  Energy 81 (2015) 125--133.

\bibitem{xsgen}
J.~Xia, A.~Scopatz,
  \href{https://doi.org/10.5281/zenodo.884490}{Flanflanagan/xsgen 0.5} (Sept
  2017).
\newline\urlprefix\url{https://doi.org/10.5281/zenodo.884490}

\bibitem{openmc}
P.~K. Romano, B.~Forget, The openmc monte carlo particle transport code, Annals
  of Nuclear Energy 51 (2013) 274--281.

\bibitem{pyne}
A.~M. Scopatz, P.~K. Romano, P.~P. Wilson, K.~D. Huff, Pyne: Python for nuclear
  engineering, Transactions of the American Nuclear Society 107 (2012) 985.

\bibitem{origen2}
A.~G. Croff, User's manual for the origen2 computer code, Tech. rep., Oak Ridge
  National Lab. (1980).

\bibitem{loop}
A.~Gul, R.~Khan, K.~Chaudri, M.~Azeem, I.~Shahzad, Validation of loop through
  pwr measured data, Annals of Nuclear Energy 111 (2018) 224--233.

\bibitem{open3bsd}
O.~S. Initiative, et~al., The bsd 3-clause license, 2013, URL
  http://opensource. org/licenses/BSD-3-Clause.Online.

\bibitem{serpentvmonte}
D.~Chersola, G.~Lomonaco, R.~Marotta, G.~Mazzini, Comparison between serpent
  and monteburns codes applied to burnup calculations of a gfr-like
  configuration, Nuclear Engineering and Design 273 (2014) 542--554.

\bibitem{ace}
J.~F. Briesmeister, et~al., MCNP--A general Monte Carlo code for neutron and
  photon transport, Los Alamos National Laboratory, 1986.

\bibitem{endf}
M.~Chadwick, P.~Oblo{\v{z}}insk{\`y}, M.~Herman, N.~Greene, R.~McKnight,
  D.~Smith, P.~Young, R.~MacFarlane, G.~Hale, S.~Frankle, et~al., Endf/b-vii.
  0: next generation evaluated nuclear data library for nuclear science and
  technology, Nuclear data sheets 107~(12) (2006) 2931--3060.

\bibitem{bondarenko}
C.~Dembia, G.~Recktenwald, M.~Deinert, Bondarenko method for obtaining group
  cross sections in a multi-region collision probability model, Progress in
  Nuclear Energy 67 (2013) 124--131.

\bibitem{weights}
D.~E. Cullen, Program groupie (version 79-1): calculation of bondarenko
  self-shielded neutron cross sections and multiband parameters from data in
  the endf/b format, Tech. rep., California Univ., Livermore (USA). Lawrence
  Livermore National Lab. (1980).

\bibitem{spectrum}
D.~G. Madland, Theory of neutron emission in fission, Tech. rep., Los Alamos
  National Lab., Theoretical Div., NM (United States) (1998).

\bibitem{linear}
M.~J. Driscoll, T.~J. Downar, E.~E. Pilat, The linear reactivity model for
  nuclear fuel management, American Nuclear Society, 1990.

\bibitem{pin-cell}
D.~E. Cullen, R.~N. Blomquist, C.~Dean, D.~Heinrichs, M.~A. Kalugin, M.~Lee,
  Y.-K. Lee, R.~MacFarlane, Y.~Nagaya, A.~Trkov, How accurately can we
  calculate thermal systems?, University of California, UCRL-TR-203892,
  Lawrence Livermore National Laboratory.

\bibitem{vision}
J.~Jacobson, A.~M. Yacout, G.~Matthern, S.~Piet, D.~E. Shropshire, C.~Laws,
  Vision: Verifiable fuel cycle simulation model, TRANSACTIONS-AMERICAN NUCLEAR
  SOCIETY 95 (2006) 157.

\bibitem{nea}
N.~E. Agency,
  \href{https://www.oecd-nea.org/ndd/reports/efc/EFC-complete.pdf}{The
  economics of the nuclear fuel cycle}.
\newline\urlprefix\url{https://www.oecd-nea.org/ndd/reports/efc/EFC-complete.pdf}

\end{thebibliography}

\end{document}